\documentstyle[prb,multicol,aps,epsf]{revtex}
\newcommand \be {\begin{equation}}
\newcommand \ee {\end{equation}}
\newcommand \eea {\end{eqnarray}}
 
 \newcommand \bi {\bibitem}

\newcommand \la {\lambda}

\input{epsf}

\begin{document}
\draft
\preprint{MA/UC3M/3/97}


\title{Langevin Dynamics of the Lebowitz-Percus Model}

\author{ F. G. Padilla(*) and F. Ritort(**)}
\address{(*) Departamento de Matem{\'a}ticas,\\
Universidad Carlos III, Butarque 15\\
Legan{\'e}s 28911, Madrid (Spain)\\
(**) Institute for Theoretical Physics,\\
University of Amsterdam\\
Valckenierstraat 67,\\
1018XE Amsterdam (The Netherlands)\\
E-Mail: padilla@dulcinea.uc3m.es,ritort@phys.uva.nl\\}
\date{\today}
\maketitle

\begin{abstract}
We revisit the hard-spheres lattice gas model in the spherical
approximation proposed by Lebowitz and Percus (J. L. Lebowitz,
J. K. Percus, Phys. Rev.{\ 144} (1966) 251). Although no disorder is
present in the model, we find that the short-range dynamical
restrictions in the model induce glassy behavior.  We examine the
off-equilibrium Langevin dynamics of this model and study the relaxation
of the density as well as the correlation, response and overlap two-time
functions. We find that the relaxation proceeds in two steps as well as
absence of anomaly in the response function. By studying the violation
of the fluctuation-dissipation ratio we conclude that the glassy
scenario of this model corresponds to the dynamics of domain growth in
phase ordering kinetics.
\end{abstract}

\section{Introduction}

The nature of slow relaxation in frustrated systems has seen a large
increase of activity in the recent years. In particular much effort has
been done in the study of relaxational dynamics in spin glasses
\cite{REV1}. These are disordered systems where slow relaxation appears
as a consequence of the frustration induced by the disorder. At low
enough temperatures the system explores a rugged free energy landscape
with high energy barriers, hence dynamics is slowed down.  But nature
offers a large variety of systems where dynamics can be exceedingly slow
in the absence of disorder. Structural glasses are examples in this
class of systems. When fast cooled, glasses do not crystallize and leave
the undercooled liquid line when the typical relaxation time exceeds the
inverse of the cooling rate. A possible scenario for the origin
of the glassy state in the absence of quenched disorder has been
recently proposed in the framework of some solvable mean-field models
\cite{BOME,MPR}. These models show the existence of three characteristic
temperatures: the melting first-order transition temperature $T_m$, a
dynamical transition temperature $T_d$ reminiscent of a spinodal
instability and a thermodynamic glass transition $T_s$ where the
configurational entropy vanishes and replica symmetry breaks 
\cite{kirk}. 
A detailed study of the dynamical equations of these mean-field models
\cite{FH} shows that these indeed correspond to the mode coupling
equations of G\"otze and collaborators for the structural glass problem
\cite{GOTZE}. One of the main results in this approximation is the
existence of a dynamical singularity ($T_d$) where dynamics is arrested
and ergodicity broken. The existence of this singularity relies on the
mean-field character of such approximation and it would be highly
desirable to understand how to include short range effects in the
theory in a systematic way.

To address this problem it is useful the study of solvable microscopic
models for a better understanding of the mechanisms
responsible for slow relaxation, the universal properties of the
off-equilibrium dynamics \cite{CUKU,FRME} as well as its connections
with the Mode Coupling approach.  In this paper we study the
off-equilibrium behavior of a solvable spherical model introduced long
time ago by Lebowitz and Percus \cite{LP}. The model has not disorder
built in and is suitable to study the role of short-range dynamical
constraints in glassy systems. We will show that the
off-equilibrium behavior of this model shares the same common features
(and in some sense, belongs to the same {\em universality} class) as
those mean-field spin glass models characterized by the absence of
anomaly in the response function. Quite remarkably the glassy scenario
of the model corresponds to the domain growth \cite{BRAY}.
We should stress also that the model introduced by Lebowitz and Percus
is not enough to provide a complete description of the glass transition
where phenomena like stretching and the characteristic two step form 
above $T_c$ in the liquid (disordered) phase are absent. In the language
of Mode Coupling theory this is due to the absence of discontinuity at
$T_c$ of the ergodicity parameter \cite{GOTZE,WHITNEY}.

But still this model deserves its interest. The new feature which this
model furnishes to the study of glassy dynamics is the effect of
dimensionality and short range dynamical constraints in the mechanisms
for glassy dynamics. This kinetic effect has been proposed in the past
as a possible explanation of the glass transition \cite{FRAN}. The
Lebowitz-Percus model (hereafter referred as LP model) incorporates the
effect of spatial correlations in the system. This allows for an
extension of the dynamical equations to include the wave vector
dependency.

The paper is divided as follows. In the section II we define the model
as well as its thermodynamic properties and write down the dynamical
equations in the Langevin approach. In section III we close the
dynamical equations for one time quantities. Section IV presents the
solution of the dynamical equations for the correlation, response and
overlap two-time functions.  Section V discusses some numerical
experiments of the model which allow for a clear identification of the
glassy scenario. Section VI discusses the nature of relaxation
processes as well as violations of the fluctuation-dissipation relation.
Finally we present the main conclusions.

\section{The Lebowitz-Percus (LP) Model}

Let us consider a lattice gas in the spherical approach. Lattice gas
models are defined on a lattice of finite dimensions where local
densities $\rho({\bf x})$ are attached to every site of the lattice. In
the hard spheres lattice gas every site of the lattice can be either
occupied
or empty. In this case the local densities $\rho({\bf x})$ can take only
the values 0 or 1. Lebowitz and Percus \cite{LP} relax these conditions on
the local densities allowing them to take any continuous value but
satisfying a global spherical constraint.  Following their work we also
introduce the additional restriction that the 
density-density correlation function between nearest neighbors
vanishes. This
restriction mimics some kind of extended hard core.  So the restrictions
on the system are,

\begin{eqnarray}
\sigma_1 \equiv \sum_{\bf x} \,\rho({\bf x})^2 -  \sum_{\bf x}
 \, \rho({\bf x}) =0 
\label{eq1}\\
\sigma_2 \equiv  \sum_{\bf x} \sum_{\bf q} \,\rho({\bf x}) \rho({\bf x+q}) =0 
\label{eq2}
\end{eqnarray}

\noindent
where ${\bf x}$ is a discrete variable that runs over the sites of a
D-dimensional lattice and ${\bf q}$ are the vectors that join a site to
its nearest neighbors. 
The thermodynamics of this
system is very well known for the general case considered by Lebowitz and
Percus \cite{LP} where a pairwise interaction potential was also
studied. For simplicity, here we will not introduce any
interaction potential between particles or additional restrictions over
the local densities.  We will see that in this case the dynamics of the
model is simply enough to be solvable displaying a non trivial
relaxational dynamics. Short range dynamical constraints are such that
relaxation turns out to be slow when a reorganization of local densities
is necessary to reach the equilibrium density of the system.  We 
start by solving the thermodynamics. After, we define the
dynamical equations of the model and find the stationary solutions of
the dynamics.

\subsection{The thermodynamics.}

In the grand canonical ensemble the partition function of the
LP model is given by,

\be
{\cal Z}_{GC}=\sum_{n=1}^\infty z^n\,Q_n(\beta)
\label{eq3}
\ee

\noindent
and $Q_n(\beta)$ is the partition function computed in the 
canonical ensemble, 

\be
Q_n(\beta)=\sum_{\lbrace\rho_l\rbrace'}\delta_
{\sigma_1,0}\delta_{\sigma_2,0}
\label{eq4a}
\ee

\noindent
where $\sigma_1$ and $\sigma_2$ have been defined previously in
eqs.(\ref{eq1},\ref{eq2}). The prime in the sum of eq.(\ref{eq4a})
indicates that the set of densities $\lbrace\rho_l\rbrace$ satisfy the
global constraint $\sum_{l=1}^N\rho_l=n$ where $n$ is the total number
of particles and $N$ the total number of sites in the lattice. Note that
the thermodynamics of this system is determined solely by the entropy
because there is no interaction energy between the particles (like in
the general case of hard spheres systems \cite{MON}). Substituting the
canonical partition function eq.(\ref{eq4a}) into the grand partition
function eq.(\ref{eq3}) we obtain in the large $n,N$ limit with $n/N$
fixed,

\be
{\cal Z}_{GC}=\sum_{\rho_l=-\infty}^{\infty}\prod_{l=1}^N
z^{\rho_l}\delta(\sigma_1)\delta(\sigma_2)
\label{eq5a}
\ee

Introducing the integral representation of the delta function we obtain,

\be {\cal Z}=\int\prod_{l=1}^N d\rho_l \int exp[\beta \sum_{l,l'} {\hat
C}(x_{l},x_{l'}) \rho_{l} (1-\rho_{l'}) + \sum_l \gamma \rho_l]
\label{eq6a}
\ee

\noindent
where $z$ is the fugacity $z=\exp(\gamma)=\exp(\mu \beta)$ and 
$\hat{C}(x_{l},x_{l'})= \lambda(x_{l}-x_{l'})$. In our case,
$\lambda(x_{l}-x_{l'})=\lambda_0$ when $x_{l}=x_{l'}$ and
$\lambda(x_{l}-x_{l'})=\lambda_1$  when $x_{l}-x_{l'}=q$, 
otherwise $\lambda(x_{l}-x_{l'})=0$.

The quadratic exponent can be diagonalized using the Fourier transform
and the partition function can be exactly evaluated in terms of the
Lagrange multipliers. The equations for the multipliers can be obtained
noting that $\frac{\partial \log {\cal Z}}{\partial \lambda_0}=\sigma_1$
and $\frac{\partial \log {\cal Z}}{\partial \lambda_1}=\sigma_2$ which
yield eq.(\ref{eq1},\ref{eq2}). Using these constraints we obtain the
following equations for the Lagrange multipliers,

\begin{eqnarray}
\frac{1}{N} \sum_k T \,\frac{\cos{({\bf k} \bf x)}} { {2 \,\tilde \lambda({\bf k})} }=
\langle \rho \rangle [ \delta( {\bf x}) -\langle \rho \rangle  ]\label{eq7a}\\
\tilde \lambda({\bf k})=\lambda_0+2 \, \lambda_1 \sum_{q} \cos({\bf k.q})
\label{eq7b} 
\end{eqnarray} 

\noindent
where ${\bf x}=0$ or ${\bf x}={\bf q}_i$ with $i=1,D$. 
In what follows we will drop
out the brackets in the notation for the average density writing $\rho$
instead of $<\rho>$. These equations can be readily solved. In the
disordered high temperature phase eq.(\ref{eq7a}) takes the simple form,

\begin{eqnarray}
\frac{1}{(2\pi)^D}\int_0^{2\pi}d{\bf k}^D \frac{T\cos({\bf k}\cdot{\bf
x})}{2\tilde{\la}}=\rho(\delta(\bf{x})-\rho)
\label{eq8c}
\\
\end{eqnarray}

\noindent
where again ${\bf x=0}$ or ${\bf x}={\bf q}_i$, $(1\le i\le D)$. A condensation phase
transition is found above two dimensions. This result can be easily
inferred after examination of eqs. (\ref{eq7a}),(\ref{eq7b}). The
phase transition corresponds to the condensation in the ${\bf
K_c}=\pi(1,1,..,1)$ direction due to the positivity of $\la_0$ and
$\la_1$.  At the transition point $\tilde \la({\bf K})=0$,
i.e. $\la_0=2\la_1 D=\la$.  The critical temperature $T_c$ and the value
of the Lagrange multiplier $\la$ are solutions to the equations,

\begin{eqnarray}
\frac{1}{(2\pi)^D}\int_0^{2\pi}d{\bf k}^D 
\frac{T_c\cos({\bf k}\cdot{\bf x})}
{2\lambda (1+\frac{1}{D} \sum_{q} \cos({\bf k.q}))}=
\rho(\delta(\bf{x})-\rho)
\label{eq8e}
\end{eqnarray}

\noindent
$x=0$ or $x={\bf q}_i$, $(1\le i\le D)$. Below the critical temperature,
condensation on the mode ${\bf K}=\pi(1,1,..,1)$ develops and the
Lagrange multipliers remain fixed to their critical values. The
condensed mass $a_{K}$ as a function of temperature satisfies the
equation,

\begin{eqnarray}
a_{K} \cos({\bf K}\cdot{\bf x})+
\frac{1}{(2\pi)^D}\int_0^{2\pi}d{\bf k}^D 
\frac{T\cos({\bf k}\cdot{\bf x})}
{2\lambda (1+\frac{1}{D} \sum_{q} \cos({\bf k.q}))}=
\rho(\delta(\bf{x})-\rho)
\label{eq8d}
\\
\end{eqnarray}

\noindent
where $x=0,{\bf q}_i$, $(1\le i\le D)$.  The mechanism for the
condensation transition in the LP model is the same as in the spherical
model of Berlin and Kac \cite{BK} and appears only for $D>D_l=2$ which
is the lower critical dimension of the model. For $D=3$ previous
equations can be solved and we find $T_c=20.158$. We should also note
that the entropy of this model diverges as $\log(T)$ at low
temperatures, hence violates the third law of thermodynamics. This
implies a finite specific heat and a finite compressibility at zero
temperature. This is not a serious drawback since this is related to the
continuous character of the local densities. To suppress this
undesirable effect one should include quantum effects \cite{theo} but
this is beyond the scope of the present paper.

\subsection {The dynamics.\label{sd1}}

The effective Hamiltonian of the LP model which takes into account 
the restrictions (\ref{eq1}) and (\ref{eq2}) imposed on the system reads,

\begin{equation}
H_{eff}(\{ {\bf x} \})=
 \lambda_0(t) \sum_x( \rho({\bf x})^2-
 \rho({\bf x}))+ \lambda_1(t) \sum_x \sum_{l=1}^{n_v} 
 \rho({\bf x})  \rho({\bf x}+{\bf q_l})
\label{eq4}
\end{equation}

\noindent
where $n_v$ is the number of 
nearest neighbors, and ${\bf q_l}$ are the vectors that join each point with 
their nearest neighbors.

We propose the following differential equation for the dynamical evolution of
the density in an open system
that can interchange particles with a reservoir,

\begin{equation}
\frac{\partial \rho({\bf x'},t) }{ \partial t }= \mu 
- \frac{\partial H_{eff}(\{ {\bf x}\},t))}{\partial  {\bf x'} }+\eta({\bf x'},t)
\label{eq5}
\end{equation}

\noindent
where $\beta$ is the inverse temperature and $\mu$ is the chemical
potential of the thermal bath. $\eta({\bf x},t)$ is a white noise
uncorrelated in time and space such that $\langle \eta({\bf x},t)
\rangle=0$ and $\langle \eta({\bf x},t) \eta({\bf x'},t') \rangle= 2 \,
T\delta({\bf x-x'}) \delta(t-t')$ where the brackets indicate realizations
over the thermal noise. From now on, we will only specify the
temporal dependence when two different times appear dropping the 
explicit time dependence for one time dynamical quantities.

We stress that in the LP model there is no energy but only entropy and
that the role of the parameters $\lambda_0(t), \lambda_1(t)$ in the
effective Hamiltonian eq.(\ref{eq4}) is to make the dynamical evolution
for the densities to fulfill the dynamical constraints
eqs(\ref{eq1},\ref{eq2}) for all times.  Note that the role of the
chemical potential $(\mu >0)$ in the dynamical equation (\ref{eq5}) is
to increase the local densities in the lattice as much as
possible. Obviously the density cannot increase indefinitely because of
the dynamical constraints. Starting from an empty lattice the dynamics
turns out to be slow when the density of the system approaches the
equilibrium density. This slowing down is a purely entropic effect and
is direct consequence of the decrease in the number of available
configurations in phase space.  In the rest of the paper and without
lost of generality, we can set the chemical potential equal to 1 in
(\ref{eq5}). The dynamical equations read,

\begin{eqnarray}
\dot{\rho}({\bf x})= 1 - \lambda_0 (2 \rho({\bf x}) - 1)
- \lambda_1 \sum_{l} (\rho({\bf x+q_l})+\rho({\bf x-q_l})) +
 \eta({\bf x})
\label{eq6}
\end{eqnarray}

\noindent
Due to the spatial translational invariance, it is easy to diagonalize 
the system using the Fourier transform, 

\begin{equation}
\tilde{\rho}({\bf k})= \frac{1}{\sqrt{N}} \sum_{ {\bf r}}
 \exp{(i {\bf k.r })} \rho({\bf r}) 
\label{eq7}
\end{equation}

\noindent
The Fourier transformed global restrictions (\ref{eq1}) and 
(\ref{eq2}) are,

\begin{eqnarray}
\sigma_1 = \sum_{ {\bf k}} | \tilde{\rho}({\bf k}) |^2
-\sqrt{N}  \tilde{\rho}({\bf 0})=0
\label{eq8a}\\
\sigma_2 = D\, \sum_{ {\bf k}} | \tilde{\rho}({\bf k}) |^2 
 \gamma(k) =0
\label{eq8}
\end{eqnarray}

\noindent
with $\gamma({\bf k})= \frac{1}{D} \sum_{l=1}^D \cos{({\bf k.q_l})}$. In terms
 of the Fourier components the effective Hamiltonian eq.(\ref{eq4}) is
 diagonal,

\begin{equation}
H( \rho({\bf k}) )= \lambda_0(t) (\sum_{ {\bf k}} | \tilde{\rho}({\bf k}) |^2
-\sqrt{N} \tilde{\rho}({\bf 0}))+2 D\, \lambda_1(t) \sum_{ {\bf k}}
 | \tilde{\rho}({\bf k}) |^2  \gamma({\bf k}) 
\label{eq9}
\end{equation} 

\noindent
and the equations of motion for the Fourier transformed densities are

\begin{eqnarray}
\dot{ \tilde{\rho}}({\bf k})= \sqrt{N} \delta({\bf k}) 
- \lambda_0 (2 \tilde{\rho}({\bf k}) - \sqrt{N} \delta({\bf k}) )
- 4 \lambda_1 D\, \tilde{\rho}({\bf k}) \gamma({\bf k}) +
 \tilde{\eta}({\bf k})
\label{eq10a}
\\
\tilde{\eta}({\bf k})= \frac{1}{\sqrt{N}} \sum_{ {\bf x}}
 \exp{(i {\bf k.x })} \eta({\bf x})  
\label{eq10}
\end{eqnarray}

This set of equations involve different uncoupled Fourier modes and can
be formally integrated for each mode like in the
Sherrington-Kirkpatrick spherical model \cite{CUDE}. Here we will follow
a different strategy and write a linear partial differential equation
for the evolution of the density. This is done in the next
section. Before we will find the stationary states of the dynamics. Due
to the absence of disorder we can also investigate the existence of
crystal states in the system. 

\subsubsection{Stationary states}

The stationary states of the model can be obtained by setting to zero
the time derivative in (\ref{eq10a}) and multiplying the equation by the 
complex conjugate of $\tilde{\rho}$. This yields,

\begin{equation}
|\tilde{\rho}_{eq}({\bf k})|^2=\frac{T+\delta({\bf k}) N \rho (\lambda_0+1)}
{2 \lambda_0+4 D \, \lambda_1 \gamma({\bf k})}
\label{eq11}
\end{equation}

\noindent
where we have used the regularization of the response function \cite{zinn},
\begin{eqnarray}
\lim_{t'\to t} <\eta(t') \rho(t)>= 2T \Theta(t-t')\\
<\eta(t') \rho(t')>= T~~~.
\label{eq110}
\end{eqnarray}
\noindent
Using that $|\tilde{\rho}({\bf 0})|^2=N \rho^2$, we note
that 
\begin{equation}
\rho=\frac{\lambda_0+1}{2 \lambda_0+4 D \, \lambda_1}+O(\frac{1}{N}).
\label{eq111}
\end{equation}
The restrictions, $\sigma_1=0$ and $\sigma_2=0$ eqs.(\ref{eq8a},\ref{eq8}) 
yield,

\begin{eqnarray}
\frac{1}{N} \sum_{\bf k} \frac{T}{2 \lambda_0+4 \lambda_1 D \gamma({\bf k})}=
\rho (1-\rho)
\label{eq12a} \\
\frac{1}{N} \sum_{\bf k} \frac{T \gamma({\bf k})}{2 \lambda_0+4 D \, \lambda_1 \gamma({\bf k})}=
-\rho^2 
\label{eq12}
\end{eqnarray}

These are the equations originally derived by Lebowitz and Percus
\cite{LP} for the thermodynamics
described in section II.A. At finite
temperature we find that the only stationary solutions are given by the
equilibrium states. This is not true at zero temperature where
ergodicity is broken. In this case we expect the
appearance of several crystalline states which nevertheless are metastable
at finite temperature. Note that the phase transition in this model is
different from the usual structural glass transformation where there is
a melting first-order phase transition from a liquid to a crystal
state. In the LP model the transition is a condensation one for $D>2$
without any latent heat. 

\subsubsection{Crystalline states}

At zero temperature, we note that there are many stationary states (its
number being proportional to the size of the system $N$). These can be
obtained by setting $T=0$ in eq.(\ref{eq10a}) which yields, after
multiplying the equation by the complex conjugate of $\tilde{\rho}$,

\be
(2 \lambda_0+4 D \, \lambda_1 \gamma({\bf k}))|\tilde{\rho}({\bf k})|^2=
N\rho\delta({\bf k})(1+\la_0)
\label{eq13c}
\ee

\noindent
For ${\bf k}=0$ this equation yields $(2\la_0+4\la_1D)\rho=(1+\la_0)$
but for ${\bf k}\neq 0$ we find different solutions depending on the
value of ${\bf k}$ and $-{\bf k}$ where the term
$(2\la_0+4\la_1D\gamma({\bf k}))$ vanishes. Then the crystal states are
characterized by a non vanishing value of $\tilde{\rho}(0)$,
$\tilde{\rho}({\bf k})$ and $\tilde{\rho}(-{\bf k})=\tilde{\rho}({\bf
k})^*$ , all the other modes being zero. The first term is the average
density of particles while the second and third ones yield the crystal
configuration. We still have to impose the conditions $\sigma_1=0$ and
$\sigma_2=0$ eqs. (\ref{eq8a}, \ref{eq8}). Then we obtain 
that $\gamma({\bf k})$
should be smaller than zero for such a solution to exist. We also obtain that
$\lambda_0=1$, and $2\,\lambda_1 \,D=-\frac{1}{\gamma({\bf k})}$ (with {\bf
k} different from zero) and the equilibrium density is given by,

\be
\rho=\frac{\gamma({\bf k})}{\gamma({\bf k})-1}
\label{eq14a}
\ee

\noindent
The number of stationary states is then proportional to the volume of
the lattice since for each value of ${\bf k}$ such that $\gamma_({\bf k})<0$
there is a stationary state. We will not extend further on
details about the crystalline ground states but limit our
discussion to the 1D case. In this case, the simplest way
to construct crystalline states is to assign a density equal to one to
each point in the lattice every $p$ sites. If $p$ is a prime
number, in the resulting periodic configuration only
$\tilde{\rho}({k=0})$, $\tilde{\rho}({k=\frac{2\,\pi}{p}})$ and
$\tilde{\rho}({k=2\,\pi(1-\frac{1}{p}}))$ are different from zero.
Then, the condition $\gamma(k)<0$ implies that $k$ lies in the interval
$(\frac{\pi}{2},\frac{3\,\pi}{2})$. So, only the states with $p=2$
(maximum filling of the lattice $\rho=1/2$) and $p=3$ (partial filling
of the lattice with $\rho=1/3$) are crystal states and fulfill the
dynamical restrictions. All other periodic states with a larger value of
$p$ have $\gamma(k)> 0$ and decay to a new $k$-state with
$\gamma(k)<0$.

\section{Dynamical solution for one time quantities.}

To study the relaxation towards the equilibrium, we will focus
our attention in the one times quantities as the density. 
We will need
them later to study the evolution of two time quantities (correlation
function, response function and the overlap among replicas). Dynamics is
such that the global constraints $\sigma_1=0$ (\ref{eq8a}) and
$\sigma_2=0$ (\ref{eq8}) are satisfied for all times, hence
$\dot{\sigma_1}=0$ and $\dot{\sigma_2}=0$.  These conditions determine
the time dependent value of the Lagrange multipliers.

\begin{eqnarray}
2\, \rho - \lambda_0+ 4\, D\, \lambda_1\, \rho -1 + 2\, R_0=0 
\label{eq13a} \\
\lambda_0 \rho- 4\, \lambda_1 \, D\, T_2+ \rho+ 2\, R_1=0
\label{eq13}
\end{eqnarray}
  
\noindent
where $\rho$ is the average density ($\rho=<\rho>$).  The quantities
$T_n$ are correlations density-density, and the $R_n$ quantities are
correlations density-noise. They are defined as follows,

\begin{eqnarray}
T_n(t)= \frac{1}{N} \sum_{\bf k} |\tilde{\rho}({\bf k})|^2 
\gamma({\bf k})^n \\
R_n(t)= \frac{1}{2N} \sum_{\bf k} (\tilde{\rho}({\bf k}) 
\tilde{\eta}^*({\bf k})+
\tilde{\eta}({\bf k}) \tilde{\rho}^*({\bf k})) 
\gamma({\bf k})^n
\label{eq14}
\end{eqnarray} 

\noindent
where $n$ is any integer number. Note that these quantities are
invariant under translations which is the main symmetry of the effective
Hamiltonian eq.(\ref{eq4}). We will see the usefulness of all these
quantities later on. We can do some remarks on the values and physical
meaning of some of them.  First of all, we note that $T_0$ is (due to
the spherical restriction) equal to the average density $\rho$. $T_1$ is
proportional to the first neighbor correlation, which is equal to zero.
$T_2$ is some kind of second neighbor correlation. This is a time
dependent quantity that needs to be calculated in order to solve the
evolution of the density. We will see that the time evolution of the
quantity $T_2$ depends on $T_3$, that $T_3$ depends on $T_4$ and so
on. In the thermodynamic limit and using the regularization of the
noise-field correlation (\ref{eq110}) we find for the $R_n$,

\begin{equation}
R_n(t) = T \; \frac{1}{N} \sum_{\bf k} \gamma({\bf k})^n 
\label{eq15}
\end{equation}

\noindent
which are time independent quantities vanishing for odd values of $n$.

We are now interested in the evolution of the average density of the
system. Taking into account that
$\langle \eta \rangle$ is of order $0(\frac{1}{\sqrt{N}})$; we get,

\begin{equation}
\dot{\rho } = 1 - \lambda_0 
(2 \rho -1) -4 D\, \lambda_1 \rho
\label{eq16}
\end{equation}

In order to solve the equations for the evolution of the density
(\ref{eq13a},\ref{eq13},\ref{eq16}),
 we need to know the dependence
of $T_2$ on time. It is easy to write the dynamical equations for the $T_n$,

\begin{equation}
\dot{T_n}=  2 \rho (1 +\lambda_0)
- 4 \lambda_0 T_n- 8 \, D\, \lambda_1 T_{n+1}+ 2 R_n
\label{eq17}
\end{equation}

As previously said, each $T_n$ depends on $T_{n+1}$.
To close this hierarchy of equations, we multiply all of them by
$ \frac{1}{n!} x^n$ and sum over $n$.
Defining $ g_T(x,t)=\sum_{n=0}^{\infty} \frac{x^n}{n!} T_n(t)$ and using that
$\frac{\partial g_T(x,t)}{\partial x }=\sum_{n=0}^{\infty} \frac{x^n}{n!} 
T_{n+1}(t)$, we get a partial differential equation for $ g_T(x,t)$,

\begin{equation}
\begin{array}{l}
\frac{\partial g_T(x,t)}{\partial t}=- 8 \, D\, \lambda_1 
\frac{\partial g_T(x,t)}{\partial x}- 4 \lambda_0 g_T(x,t)+
2 \rho (1 + \lambda_0) e^x
+ 2 g_N(x) 
\\
 g_N (x) = \sum_n \frac{1}{n!} x^n R_n = \frac{T}{N} \sum_{\bf k}
\exp{(x\, \gamma({\bf k}))}
\end{array}
\label{eq18}
\end{equation}

We can formally integrate this partial differential equation using
the method of the characteristic curves,

\begin{equation}
\begin{array}{c}
g_T(t,x)= g_0(x-\int_0^t 8\,D \lambda_1 dt') \exp{(-\int_0^t 4 \lambda_0 dt')}+
\nonumber
\\
\int_0^t [ 2 \rho (1 + \lambda_0) 
\exp{(x-\int_{t'}^t 8\,D\, \lambda_1 dt'')}+ 
2 g_N(x-\int_{t'}^t 8\,D\, \lambda_1 dt'')] 
\exp{(-\int_{t'}^{t} 4 \lambda_0 dt'')} dt'
\end{array}
\label{eq19}
\end{equation}

\noindent
where the $g_0$ function is determined by the initial condition,
$g_T(0,x)=g_0(x)$. Once the value of $g_T(t,x)$ is known, we can
obtain the different elements of the hierarchy by taking derivatives 
with respect to $x$, 
$T_n(x,t)=\frac{\partial^n g_T(x,t)}{\partial x^n}|_{x=0}$.

\section{The hierarchy of equations for two time quantities.}

In this section, we write the dynamical equations for the correlation,
response and overlap functions. These are defined by,

\begin{eqnarray}
G(t,t',{\bf x})=\frac{1}{N} \sum_{\bf x'} \rho({\bf x'}+{\bf x},t) \eta({\bf x'},t')=
\frac{1}{N} \sum_{\bf k} \tilde{\rho}({\bf k},t) \tilde{\eta}(-{\bf k},t')
\exp{(i {\bf k.x})}
\label{eq20} \\
C(t,t',{\bf x})=\frac{1}{N} \sum_{\bf x'} \rho({\bf x'}+{\bf x},t) \rho({\bf x'},t')=
\frac{1}{N} \sum_{\bf k} \tilde{\rho}({\bf k},t) \tilde{\rho}(-{\bf k},t')
\exp{(i {\bf k.x})}
\label{eq20a}  \\
Q(t,t')=\frac{1}{N} \sum_{\bf x'} \rho_1({\bf x'}+{\bf x},t) \rho_2({\bf x'},t)=
\frac{1}{N} \sum_{\bf k} \tilde{\rho}_1({\bf k},t) \tilde{\rho}_2(-{\bf k},t)
\label{eq20b}
\end{eqnarray}

$C(t,t')$ (\ref{eq20a}) is the density-density
correlation function and measures how fast the configurations decorrelate in
time. The response function (\ref{eq20}) measures the change of the local
density in a point of the lattice at time $t$ when the chemical
potential is locally changed in another point of the lattice at distance
$x$ at time $t'$ \cite{texte1}. Finally the overlap function
(\ref{eq20b}) measures how fast two different copies of the system
(initially in the same configuration at $t'$,
i.e. $\rho_1(x,t)=\rho_2(x,t')$) decorrelate in time when submitted to
different realizations of the noise for $t>t'$ \cite{CUDE,BBM}.  From
now on and in the rest of the paper we will take the convention $t >
t'$. To obtain their time evolution we proceed as in the case of the
density and define hierarchies for two time quantities as follows,

\begin{eqnarray}
G_n(t,t',{\bf x})=
\frac{1}{N} \sum_{\bf k} \tilde{\rho}({\bf k},t) \tilde{\eta}(-{\bf k},t')
\gamma({\bf k})^n \cos{({\bf k.x})} 
\label{eq21}\\
C_n(t,t',{\bf x})=\frac{1}{N} 
 \sum_{\bf k} \tilde{\rho}({\bf k},t) \tilde{\rho}(-{\bf k},t')
\gamma({\bf k})^n \cos{({\bf k.x})} 
\label{eq21a} \\
Q_n(t,t')=\frac{1}{N} 
 \sum_{\bf k} \tilde{\rho}_1({\bf k},t) \tilde{\rho}_2(-{\bf k},t)
\gamma({\bf k})^n
\label{eq21b}
\end{eqnarray}

Using (\ref{eq10a}) we get the following equations, 

\begin{eqnarray}
\frac{\partial G_n}{\partial t}(t,t',{\bf x})=
-4\,D\, \lambda_1(t)  G_{n+1}(t,t',{\bf x}) -2 \lambda_0(t) G_n (t,t',{\bf x}) 
\label{eq220} \\
\frac{\partial C_n}{\partial t}(t,t',{\bf x})=
-4\,D\, \lambda_1(t)  C_{n+1}(t,t',{\bf x}) -2 \lambda_0(t) 
C_n (t,t',{\bf x})+ \rho(t') (1+\lambda_0(t)) 
\label{eq221} \\
\frac{\partial Q_n}{\partial t}(t,t')=
-8\,D\, \lambda_1(t) Q_{n+1}(t,t') -4 \lambda_0(t) 
Q_n (t,t') + 2 \rho(t) (1+\lambda_0(t))
\label{eq222}
\end{eqnarray}

\noindent 
For the response function we have $<\rho(t') \eta(t)>=0$.
Defining  the following generating functions,

\begin{eqnarray}
\Gamma_G(t,t',{\bf x},y)=\sum_{n=0}^{\infty} \frac{y^n}{n!} G_n(t,t',{\bf x})
\label{eq23a} 
\\
\Gamma_C(t,t',{\bf x},y)=\sum_{n=0}^{\infty} \frac{y^n}{n!} C_n(t,t',{\bf x})
\label{eq23b} 
\\
\Gamma_Q(t,t',y)=\sum_{n=0}^{\infty} \frac{y^n}{n!} Q_n(t,t')
\label{eq23}
\end{eqnarray}

\noindent
we can close the hierarchies (\ref{eq23a}),(\ref{eq23b}) 
and (\ref{eq23}),

\begin{eqnarray}
\frac{\partial \Gamma_G }{\partial t}(t,t',{\bf x},y)=
-4\,D\, \lambda_1(t) \frac{\partial \Gamma_G}{\partial y}(t,t',{\bf x},y) 
-2 \lambda_0(t) \Gamma_G(t,t',{\bf x},y) 
\label{eq24}\\
\frac{\partial \Gamma_C}{\partial t}(t,t',{\bf x},y)=
-4\,D\, \lambda_1(t) \frac{\partial \Gamma_C}{\partial y}(t,t',{\bf x},y)
 -2 \lambda_0(t) \Gamma_C (t,t',{\bf x},y)+ \rho(t') (1+\lambda_0(t)) e^y 
\label{eq24a}\\
\frac{\partial \Gamma_Q}{\partial t}(t,t',y)=
-8\,D\, \lambda_1(t) \frac{\partial \Gamma_Q}{\partial y}(t,t',y)
 -4 \lambda_0(t) \Gamma_Q(t,t',y) + 2 \rho(t) (1+\lambda_0(t))  e^y
\label{eq24b}
\end{eqnarray}

The initial condition in
equations (\ref{eq21},\ref{eq21a},\ref{eq21b}) is obtained setting $t=t'$.
Using the regularization of the
field-noise correlation at $t=t'$ (\ref{eq110}), the initial 
condition for the response function $\Gamma_G^{(0)}(t',{\bf x},y)$ is given by

\begin{eqnarray}
\Gamma_{G}^{(0)}(t',{\bf x},y)=\Gamma_G(t',t',{\bf x},y)=\sum_{n=0}^{\infty} \frac{y^n}{n!} 
\frac{T}{N} \sum_{\bf k} \gamma(k)^n \cos{({\bf k.x})} 
\label{eq241}
\end{eqnarray}

\noindent
For the correlation function, we have,

\begin{eqnarray}
\Gamma_{C}^{(0)}(t',{\bf x},y)=\Gamma_C(t',t',{\bf x},y)=\sum_{n=0}^{\infty}
\frac{y^n}{n!}  \frac{1}{N} \sum_{\bf k} |\tilde{\rho}({\bf k},t')|^2
\gamma({\bf k})^n \cos{({\bf k.x})}
\label{eq242}
\end{eqnarray}

\noindent
In particular, for ${\bf x}=0$ we have $\Gamma_{C}^{(0)}(t',0,y)=g_T(t',y)$. So
we must use the generating function for the density as the initial
condition for the generating function of the correlation.

\noindent
The same initial condition needs to be used for the generating function
for the overlap between replicas,
 
\begin{eqnarray}
\Gamma_{Q}^{(0)}(t',y)=\Gamma_Q(t',t',y)=g_T(t',y)
\label{eq243}
\end{eqnarray}

Equations (\ref{eq24},\ref{eq24a},\ref{eq24b}) with their initial
conditions (\ref{eq241},\ref{eq242},\ref{eq243}) can be formally 
solved as we did for the generating function 
for the hierarchy of the density (\ref{eq18}). 

\subsubsection{The equilibrium solution.}

Using the integral expressions for $\Gamma_C$, $\Gamma_G$ and  
$\Gamma_Q$, we can find the equilibrium values for the two
time quantities.
In equilibrium, $\rho$, $\lambda_0$ and $\lambda_1$ are time
independent, and the integral expressions can be simplified.
Imposing also the equilibrium solutions as initial condition 
and using equation (\ref{eq11}) we get the following results,

\begin{eqnarray}
\Gamma_C^{eq}(t,t',{\bf x},y)=\frac{1}{N} \sum_{\bf k} \frac{T \cos{({\bf k.x})}
\exp{((y-4 \,D \, \lambda_1 (t-t')) \gamma({\bf k})-2 \, \lambda_0 (t-t'))}}
{2 \, \lambda_0+ 4 \,D \, \lambda_1 \gamma({\bf k}) }+ \rho^2 \exp{(y)}
\label{eq251}\\
\Gamma_G^{eq}(t,t',{\bf x},y)=\frac{1}{N} \sum_k T \cos{({\bf k.x})}
\exp{((y-4 \,D \, \lambda_1 (t-t')) \gamma({\bf k})-2 \, \lambda_0 (t-t'))}
\label{eq252}\\
\Gamma_Q^{eq}(t,t',{\bf x},y)=\frac{1}{N} \sum_{\bf k} \frac{T \cos{({\bf k.x})}
\exp{((y-8 \,D \, \lambda_1 (t-t')) \gamma({\bf k})-4 \, \lambda_0 (t-t'))}}
{2 \, \lambda_0+ 4 \,D \, \lambda_1 \gamma({\bf k}) }+ \rho^2 \exp{(y)}
\label{eq253}
\end{eqnarray}

It can be readily seen that the equilibrium solution is time
translational invariant and that the Fluctuation Dissipation Theorem
(hereafter denoted as FDT) for the generating function is satisfied,
i.e.  $\frac{\partial \Gamma_C(t,t',{\bf x},y)}{\partial t'}= \Gamma_G(t,t',{\bf x},y)
$. This implies the validity of the FDT for any of the elements of the
hierarchy of equations. In particular, for the usual response and
correlation functions we get $\frac{\partial C_0(t,t',{\bf x})}{\partial t'}=
G_0(t,t',{\bf x}) $. Also we note that $\Gamma_C^{eq}(2(t-t'),{\bf x},y)=
\Gamma_Q^{eq}(t-t',{\bf x},y)$ as expected \cite{BBM}. 

In the thermodynamic limit the previous discrete sums in 
eqs.(\ref{eq251},\ref{eq252},\ref{eq253}) become integrals. In the
condensed phase we obtain the following expressions,

\begin{equation}
\begin{array}{c}
\Gamma_C^{eq}(t-t',{\bf x},y)=a_{\bf K}\cos({\bf K}\cdot{\bf x}) \exp{(y)}+
\nonumber
\\
\frac{1}{(2\pi)^D}\int_0^{2\pi}d{\bf k}^D
\frac{T\cos({\bf k}\cdot{\bf x}) \,
\exp{((y-4 \,D \, \lambda_1 (t-t')) \gamma({\bf k})-2 \, \lambda_0 (t-t'))}}
{2\lambda (1+\frac{1}{D} \sum_{\bf g} \cos({\bf k.q}))}-\rho^2 \exp{(y)}
\\
\Gamma_G^{eq}(t-t',{\bf x},y)=a_{\bf K}\cos({\bf K}\cdot{\bf x}) \exp{(y)}+
\nonumber
\\
\frac{1}{(2\pi)^D}\int_0^{2\pi}d{\bf k}^D
T\cos({\bf k}\cdot{\bf x}) \,
\exp{((y-4 \,D \, \lambda_1 (t-t')) \gamma({\bf k})-2 \, \lambda_0 (t-t'))}
\\
\Gamma_Q^{eq}(t-t',y)=a_{{\bf K}} \exp{(y)}+
\nonumber
\\
\frac{1}{(2\pi)^D}\int_0^{2\pi}d{\bf k}^D
\frac{T \,
\exp{((y-8 \,D \, \lambda_1 (t-t')) \gamma({\bf k})-4 \, \lambda_0 (t-t'))}}
{2\lambda (1+\frac{1}{D} \sum_{\bf q} \cos({\bf k.q}))}-\rho^2 \exp{(y)}
\end{array}
\label{eq254}
\end{equation}

\noindent
where the values of $\lambda$ and $a_{\bf K}$ are determined by equations
(\ref{eq8d}). Above the critical temperature in the disordered phase the
expressions for $\Gamma_C$, $\Gamma_G$ and $\Gamma_Q$ are very similar to
the previous ones except for the Lagrange multipliers which do not verify
$\lambda_0=2 \,D\, \lambda_1$ and $a_{\bf K}=0$. In this case the Lagrange
multipliers are determined by equations (\ref{eq8c}).

\section{Dynamical solution}

\subsection{General Method}

The solution to the previous equations is quite involved in the
off-equilibrium regime and cannot be exactly calculated even if some
results can be obtained in the asymptotic long time limit. While it is
possible to simplify the analytical expressions using Laplace
transformations the analytical analysis of the dynamical equations
appears to be tedious. Here we follow a different and more
straightforward strategy and numerically investigate the solution to the
dynamical equations.  To understand the nature of the off-equilibrium
behavior of the model we have numerically integrated the dynamical
equations by truncating the hierarchy up to a given finite number of
elements.  We have investigated the cases $D=$1 and 3. In the first case,
there is no phase transition while there is a transition in the second
one. For the numerical integration of the equations we have considered
between 100 and 500 elements of the hierarchy for $D=1$ and between 5000
and 20000 elements in the condensed phase in $D=3$ (where relaxation is
slower).  We note that all figures in this section when plotted in a
logarithmic scale are always in base $\log_{10}$. In all the cases, the
integration was performed with an Euler method of second order. Some of
this results were tested with a fourth order Euler method.

\subsection{One time quantities}

\subsubsection{Relaxation of the density}

The first behavior we can study is the relaxation of the density to its
equilibrium value. In one dimension we find that the relaxation is
exponential with time for $T$ different from zero as expected in the
disordered phase. For $T=0$, we find different behaviors depending on
the initial condition as found in the spherical Sherrington-Kirkpatrick
model \cite{CUDE}. If the initial condition has a macroscopic projection
on the equilibrium state then the relaxation is exponential. If the
projection on the equilibrium state is zero, then the density of the
system relaxes to the equilibrium with a power law, $t^{-1}$.

In three dimensions, starting from a non-equilibrium initial condition
the system relaxes exponentially fast above the critical temperature in
the disordered phase. In the condensed phase the relaxation is algebraic
with a $t^{-1}$ power law (figure 1). The behavior of the density in the
LP model is very similar to the relaxation of the energy in the
disordered spherical SK model \cite{CUDE}.

\begin{figure}
\begin{center}
\leavevmode
\epsfysize=230pt{\epsffile{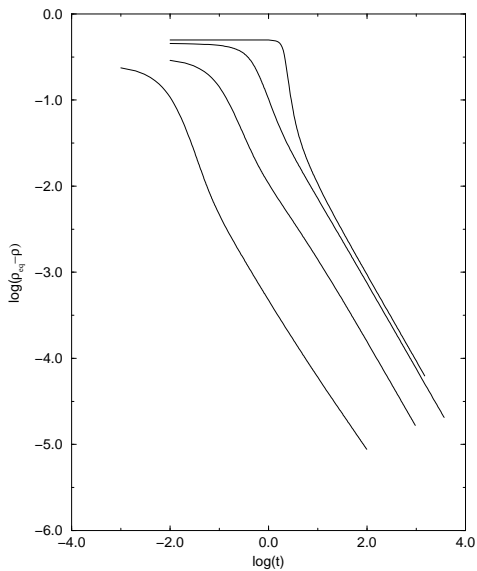}}
\end{center}
  \protect\caption[1]{
Time dependent density in 3D $\rho_{eq}-\rho$ (from top to bottom) $T=0.0001$,
0.1,1 and 10, with homogeneous initial condition.
The equilibrium densities at  $T=0.0001$ is 0.499950, at  $T=0.1$ 
is 0.454950, at  $T=1.$ is 0.309017 and at  $T=10.$ is 0.256246.
    \protect\label{FIG1}
  }
\end{figure}

\subsubsection{Hysteresis effects.}

We have performed some temperature cycling experiments in our system.
Starting from a random high temperature configuration (for $D=3$ we
start above the condensation transition temperature) we let the system
equilibrate at this temperature and later on we decrease the temperature
at a finite rate. As far as we are mainly interested in the slow 
cooling regime, to integrate the equations we change the 
temperature in the differential equations on constant steps of 
$\Delta T$. The cooling rate is given by $r=\frac{\Delta T}{t^*}$,
where $t^*$ is the time the system spends in a given constant 
temperature.  We observe that the system departs from the
equilibrium line at a temperature $T^*$ which decreases as the cooling
rate decreases. The inverse of the cooling rate yields an estimate of
the relaxation time at that temperature $T^*$ . Below $T^*$ the system
fails to relax to the equilibrium and remains at a density lower than
the equilibrium value. Once we reach zero temperature, we start to
increase it at the same rate. We find that the non-equilibrium line
crosses the equilibrium one which indicates that the system also fails
to relax to the equilibrium but now it remains at a density higher than
the equilibrium value. Finally, the non equilibrium line merges again
the equilibrium line at a temperature of the same order of $T^*$.  This
behavior can be observed in 1-D (figure 2) as well as in 3-D (figure
3).  In these experiments there is no apparent difference between the
3-D and the 1-D case indicating that this is a general non-equilibrium
effect unsensitive to the existence of a phase transition. Similar
effects are observed in the 1-D Ising model \cite{1D} as well as in the
Backgammon model \cite{BG1}.

\begin{figure}
\begin{center}
\leavevmode
\epsfysize=230pt{\epsffile{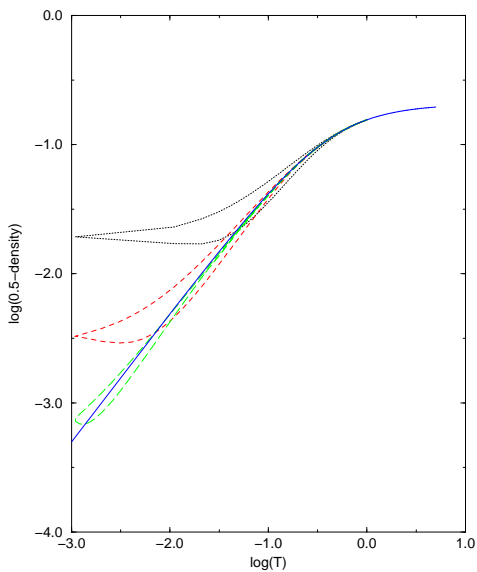}}
\end{center}
  \protect\caption[2]{
Cooling experiment in 1-D. 0.5-density versus temperature.
We show the equilibrium value for density (continuous line) and the
value for the density at different cooling rates. These are
(from top to bottom) 0.1,0.01 and 0.001. 
    \protect\label{FIG2}
  }
\end{figure}

\begin{figure}
\begin{center}
\leavevmode
\epsfysize=230pt{\epsffile{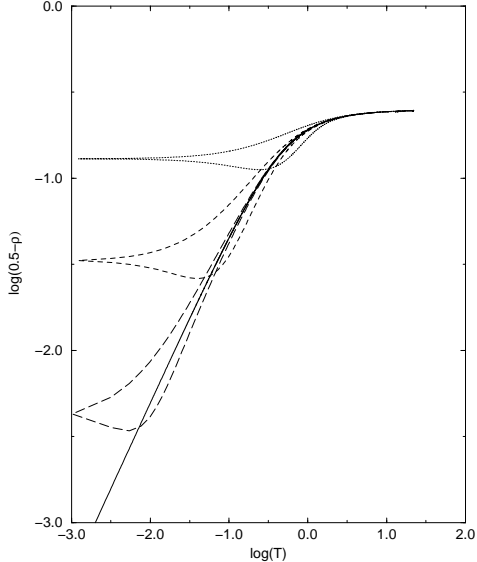}}
\end{center}
  \protect\caption[3]{
Cooling experiment in 3-D. 0.5-density versus temperature.
We show the equilibrium value for density (continuous line) and the
value for the density at different cooling rates. These are
(from top to bottom) 0.1,0.01 and 0.001.
    \protect\label{FIG3}
  }
\end{figure}

\subsubsection{Heating Experiments}

An striking feature of this system is that although no potential energy
has been introduced there appear to be a large number of stable
crystalline states at zero temperature which have an infinite lifetime.
To see the effect of these crystalline states on the dynamics we have done
some heating experiments. At $T=0$ an in 1-D, we have
constructed different periodic crystalline initial conditions (putting
the density every $p$ sites equal to one).  As we have shown in
section (\ref{sd1}), crystalline configurations with $p=2,3$ are
metastable states at zero temperature. Numerically we find that the
configurations where $p=4$ and $p=6$ decay to the configuration $p=2$,
and that $p=5$ and $p=7$ does not decay to $p=2$ or $p=3$, but to
solutions with densities between $\frac{1}{2}$ ($p=2$) and $\frac{1}{3}$
($p=3$). These results are simply explained noting that, although the
restriction $\sigma_2=0$ is a global one, apparently the only way to
fulfill this restriction is imposing that every occupied site is
surrounded by empty sites.

At finite temperature, the metastable states have a finite lifetime and
decay to the equilibrium state. However, at low temperatures, the
lifetime of metastable states can be very large and their effects on the
dynamics can then be observed. Suppose we prepare the system in such a
way that the initial condition belongs to the basin of attraction of one
of these stationary states. If we perform a heating experiment
in which we raise the temperature of the system at a finite rate, then 
we find that the system is trapped for some time near the basin of attraction.
In figure 4, we show the case where we start from a non metastable state
with $p=5$. We see that first it decays to the crystal state (the first
plateau seen in the figure) an later on it relaxes to the equilibrium line.

\begin{figure}
\begin{center}
\leavevmode
\epsfysize=230pt{\epsffile{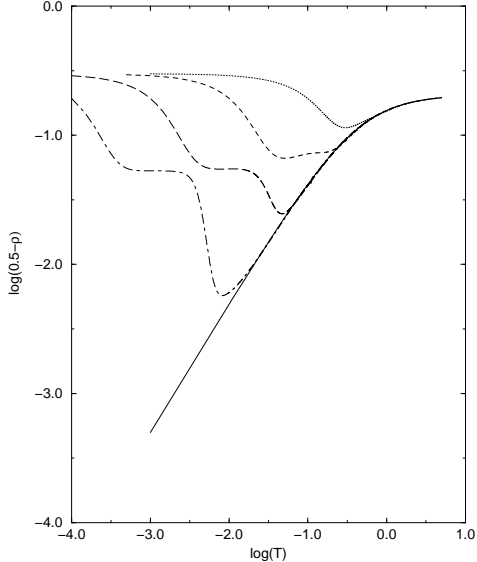}}
\end{center}
  \protect\caption[4]{
Heating experiment in 1-D. 0.5-density versus temperature.
Starting from a 5 period 1-D crystal initial condition.
We show the equilibrium value for density (continuous line) and the
value for the density at different cooling rates. These are
(from top to bottom) 0.1,0.01, 0.001 and 0.0001.
    \protect\label{FIG4}
  }
\end{figure}

\subsubsection{Kinetic growth.}

In the LP model, contrarily to the case of the spherical
Sherrington-Kirkpatrick model \cite{Kos}, it is possible to investigate
kinetic growth phenomena, i.e. how the condensed domains grow as time
goes by. This is interesting because allows to ascertain how relevant is
dimensionality in non equilibrium phenomena. Let us consider the
densities in Fourier space eq.(\ref{eq7}). Using equation (\ref{eq10a})
we obtain its time evolution,

\begin{equation}
\frac{1}{2} \frac{d\, |\tilde{\rho}({\bf k})|^2}{d\,t}
=  N \delta(k) \rho 
- \lambda_0 (2 |\tilde{\rho}({\bf k})|^2 - N \delta(k) \rho )
- 4 \lambda_1 D\,|\tilde{\rho}({\bf k})|^2  \gamma(k) + T
\end{equation}

Integrating this equation with the values for the Lagrange multipliers
obtained from the hierarchy of the density, we get the Fourier
transformed densities for the infinite system.  During the evolution
from a non-equilibrium initial condition, we see how a peak around
$k=\pi$ is formed (in 3-D, around ${\bf k}=(\pi,\pi,\pi)$). As
temperature decreases the peak becomes sharper. For a better
understanding of the results, we can appeal to the equilibrium
relations. Using (\ref{eq11}), (\ref{eq12a}) and (\ref{eq12}) we can
calculate an effective temperature for any $|\tilde{\rho}({\bf
k})|^2$. Obviously, in equilibrium, all of them will have the same
effective temperature, the equilibrium one. 

In figure 5, we show the values of the effective temperature for values
of ${\bf k}=k\,(1,1,1)$, with $k$ ranging from 0 to $\pi$ in the 3-D
case (the figure is symmetric around $k=\pi$). For times larger than a
typical time $t^*\simeq 1$ a plateau appears in the effective
temperature and its value decays very fast to $T$, the temperature of
the heat bath.  For values of $k$ ranging from zero up
to a given value $k_{max}$, all the modes have approximately the same
effective temperature which is the equilibrium one
\cite{texte2}. There is a given value of $k$ (let us call it $k^*$)
 where the effective temperature becomes
infinite and above this value of $k^{max}$ the effective temperature is
not defined (This means that there does not exist an equivalent
equilibrium system described by the dynamical densities
$|\tilde{\rho}({\bf k})|^2$).

\begin{figure}
\begin{center}
\leavevmode
\epsfysize=230pt{\epsffile{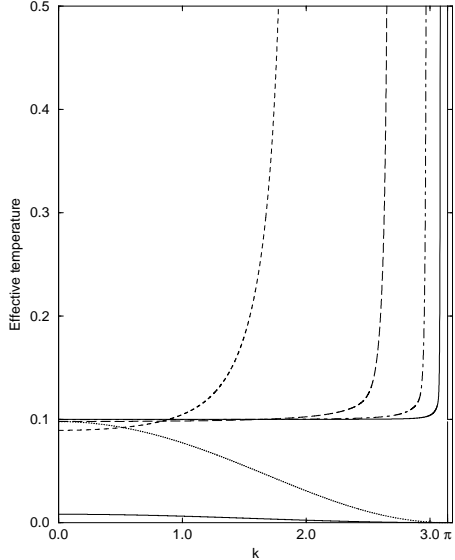}}
\end{center}
  \protect\caption[5]{
Effective temperatures versus k in 3-D and
at $T=0.1$. It is shown for different values of time:
 0.01,0.1,1,10,100,1000. The vertical line on the right of the plot 
 is $k=\pi$.
  }
\end{figure}

With the value of $k_{max}$ we can also estimate the correlation
length using the relation $\xi\simeq \frac{1}{\pi-k_{max}}$. We find
that the correlation length diverges as $t^{\frac{1}{2}}$ typical of
domain growth in ferromagnets with non conserved dynamics (figure 6).

The dynamics in the LP model can be intuitively understood in the
framework of the kinetic growth scenario \cite{BRAY}. For times less
than $t^*$ there is no plateau in the effective temperature because the
system is nearly empty and the system is filling the lattice in a random
and uncorrelated way. As soon as the density becomes large enough, any
change of the density in a site of the lattice is correlated with that
of the nearest neighbors and any change in the density requires a
reorganization of the local densities inside a given domain.  So, a
critical and finite time $t^*$ is needed till the first domains
appears. At times larger than $t^*$ correlated domains start to grow in
time. At length scales smaller than the growing correlation length $\xi$
the system is in local equilibrium (the effective temperature is the
equilibrium temperature) while it is completely disordered (the
effective temperature is infinite) at larger length scales. The domain
growth scenario is nicely reproduced in the LP model.

\begin{figure}
\begin{center}
\leavevmode
\epsfysize=230pt{\epsffile{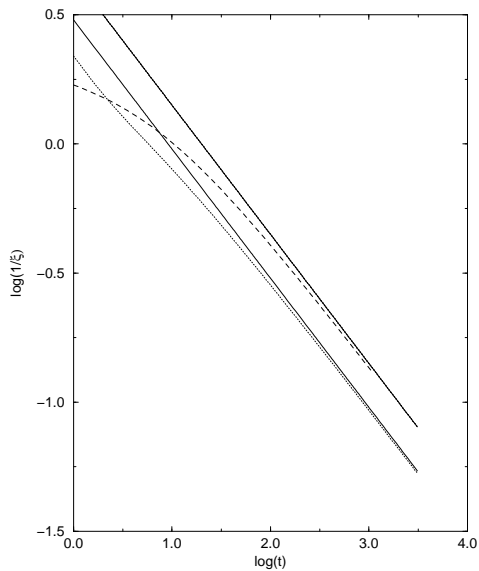}}
\end{center}
  \protect\caption[6]{ Inverse of the correlation length, $1/\xi$,
against time in 3-D and at $T=0.1$ (dotted line) and 1 (dashed line).
The straight lines show $t^{-1/2}$ behavior.  }
\end{figure}

\subsection{Two Times Quantities.}

\subsubsection{The Correlation and Response Functions.}

In this section, we are going to study the correlation and response
functions  in equations (\ref{eq221}) and (\ref{eq222}) with $x=0$.
We consider the following normalized correlation function,

\begin{equation}
C_{norm}(t,t')=\frac{C_0(t,t',x=0)-\rho_{eq}^2}
{C_0(t',t',x=0)-\rho_{eq}^2}=
\frac{C_0(t,t',x=0)-\rho_{eq}^2}{\rho_0-\rho_{eq}^2}
\end{equation}

\noindent
where $\rho_0$ is the initial density at time $t'$ and $\rho_{eq}$
is the equilibrium density corresponding to the system
at temperature $T$. In what follows we redefine the time variables $t'=t_w$
and consider the behavior of correlation functions for different values
of $t_w$.

Our results confirm the simplest Mode Coupling scenario where relaxation
proceeds in two steps, the $\alpha$ and $\beta$ relaxation processes.  A
comment now seems to be appropriate. Originally Mode Coupling theory was
devised to understand the equilibrium dynamics of liquids in the
vicinity of the glass transitions. G\"otze and collaborators
\cite{GOTZE} have proposed a general mathematical framework for the
understanding of equilibrium relaxational processes which take place in
the vicinity of a bifurcation instability. The model we propose here is
a very simplified version of the glass scenario where there is no
discontinuity of the ergodicity parameter in the bifurcation point
(i.e. at the condensation phase transition temperature). In fact, above
$D_l=2$ the LP model exhibits a usual second order phase transition with
classical critical exponents $\nu=1/(D-2), \eta=0$ ($D\le 4$). The
dynamical critical exponent is $z=2$ typical of mean-field theory. This
implies that the relaxation time diverges close to the transition
temperature $T_c$ like $\tau \sim (T-T_c)^{-2/(D-2)}$. In the critical
point equilibrium time correlations decay like $t^{-(D-2)/z}$, i.e. like
$t^{-1/2}$ for $D=3$. Among other results these scaling relations led
naturally to the $1/t$ decay found in figure 1 for the density in all
dimensions at the critical point and below it. Above $T_c$ the scaling
behavior $C(t)\sim t^{-(D-2)/z}f(t/\tau)$ holds where $\tau$ is the
divergent characteristic time scale and the superposition principle is
then valid. We want to stress that, contrarily to other scenarios for
glassy relaxations (see the discussion by G\"otze\cite{WHITNEY}) in
this case there is only one relaxation process above $T_c$ since there
is not discontinuity in the ergodicity parameter at $T_c$. But still it
is quite interesting to investigate the extension of the Mode Coupling
scenario to the off-equilibrium dynamics below $T_c$. In this case, it
is necessary to consider the initial time dependence in the dynamical
equations. This implies the emergence, among others, of new off
equilibrium phenomena like aging, i.e. the dependence of the evolution
of the system on the initial time state. Also below $T_c$
off-equilibrium correlation functions are expected to display the
characteristic two steps form since the ergodicity parameter (i.e. the
Edwards-Anderson parameter) is already finite.

We will consider $t_w$ larger than $t^*$, i.e. the typical time needed
to reach a macroscopic density in the lattice starting from a nearly
empty lattice. For values of $t_w<t^*$, the system is quite far from the
asymptotic long time regime and obviously the two step form is hardly
seen.  There are qualitative differences between the 1-D and 3-D cases
since in 3-D there is a condensed phase while the system is always
disordered in the 1-D case. The off-equilibrium correlation function in
3-D decays in two steps: the $\beta$ process or stationary part with a
decay to a plateau with a $t^{-1/2}$ and the
slow $\alpha$ process where the density-density
correlation function decays like $t^{-\frac{3}{2}}$. In the $\beta$
process the relaxation is stationary and depends only on the time
difference $t-t_w$ while in the slow $\alpha$ process there is aging and
the correlation functions depend on both $t$ and $t_w$ (figure 7). The
scaling behavior $t/t_w$ is well reproduced in the off-equilibrium
regime (figure 8). In 1-D the behavior is similar except for the
plateau which is absent (only a small inflexion for the correlation
function at short times is observed). In 3-D the plateau
persists over an arbitrarily long time scale which grows with $t_w$
whereas for 1-D  aging is interrupted and disappears
when $t_w$ is of the order of the finite relaxation time.

\begin{figure}
\begin{center}
\leavevmode
\epsfysize=230pt{\epsffile{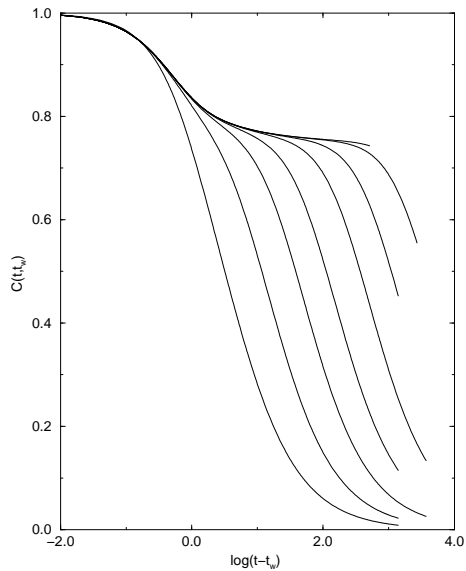}}
\end{center}
  \protect\caption[7]{ 
Correlation function versus $t-t_w$ in
3-D at $T=0.1$. Different waiting times are shown.
From top to bottom $t_w=10000$,1000,300,100,30,10,3,1.
 }
\end{figure}

\begin{figure}
\begin{center}
\leavevmode
\epsfysize=230pt{\epsffile{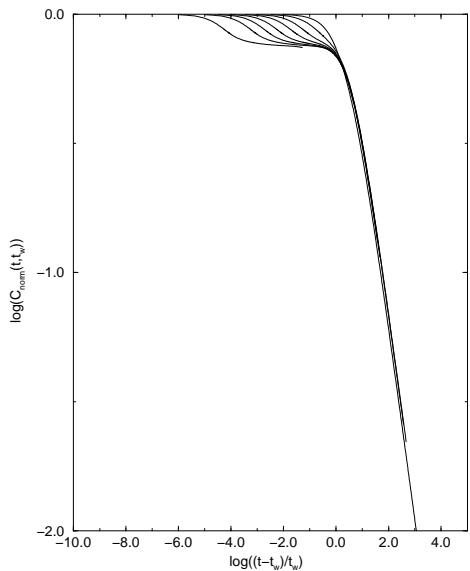}}
\end{center}
  \protect\caption[8]{
Correlation function versus $(t-t_w)/t_w$ in
3-D and T=0.1. Different waiting times are shown,
 $t_w=1000$,300,100,30,10,3,1. For times larger than 1, all of them merge in
one curve except for $t_w=1$ (the lines on the left) which is too short.
}
\end{figure}

The dynamical scenario in the LP model can be inferred from a study of
the response function. In any dimension the response function decays
very fast to zero showing no aging in the asymptotic long time
regime. Indeed, in 3D we see that $G(t,t_w)$ decays like $t^{-3/2}$
for long times and then the integrated response function decays as 
$t^{-1/2}$. This is an indication of the short-time memory of the
system. In the context of glassy dynamics this indicates that the LP
model has no anomaly in the response function \cite{BBM}. 

\begin{figure}
\begin{center}
\leavevmode
\epsfysize=230pt{\epsffile{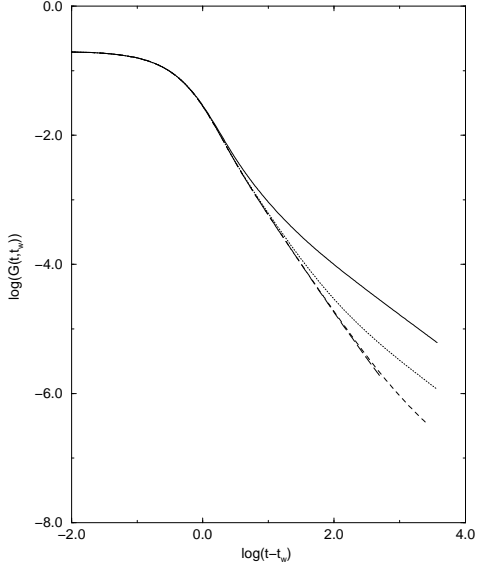}}
\end{center}
  \protect\caption[9]{
Response function, $G(t,t_w)$, versus $t-t_w$ in 3-D
and for $T=0.1$.  Different waiting times are shown.
From top to bottom $t_w=10$,100,1000,10000.
}
\end{figure}

\subsection{The Fluctuation Dissipation Theorem.}

Once we know the values of the correlation and response functions, we
can study the fluctuation-dissipation ratio,

\begin{equation}
X(t,t_w)=\frac{G_0(t,t_w)}{\frac{\partial C_0(t,t_w)}{\partial t_w}}~~~.
\end{equation}

This ratio is equal to 1 in equilibrium (recall that, by definition, a
factor $T$ has been absorbed in the response function). Roughly, we find
that $X(t,t_w)$ is approximately equal to 1 for $t-t_w < t_w$, showing
that the system is in local equilibrium in the stationary regime. For
times $t-t_w$ larger than $t_w$ the $X$ decays to zero very fast.
 We find the same qualitative behavior in 1-D and 3-D. This
is expected in the absence of anomaly since the response function decays
very fast to zero in this regime. If we look more carefully (see figure
10), we find that there are two different regimes.  For small values of
$t_w$ ($t_w<t^*\simeq 1$ when the lattice starts to be filled of
particles), the value of $X$ decreases monotonically; but for larger
values of $t_w$, $X$ increases with $t$ until it reaches a maximum and
after decreases very fast. Numerically, we can
extrapolate that for $t_w$ tending to infinity and $t-t_w<t_w$ (i.e. in
the $\beta$-regime) the $1-X(t,t_w)$ tends to zero as $(t-t_w)^{-1}$.

Using the fact that for long times the response functions decays
as $t^{-3/2}$ and that the  $X(t,t_w) \simeq 1$, we can conclude
that the correlation function decays to the plateau with a power
law $t^{-1/2}$. This result has been checked directly by fitting
the decay of the correlation function to the plateau (see below for
an estimate of the value of the plateau) for large values of $t_w$.

\begin{figure}
\begin{center}
\leavevmode
\epsfysize=230pt{\epsffile{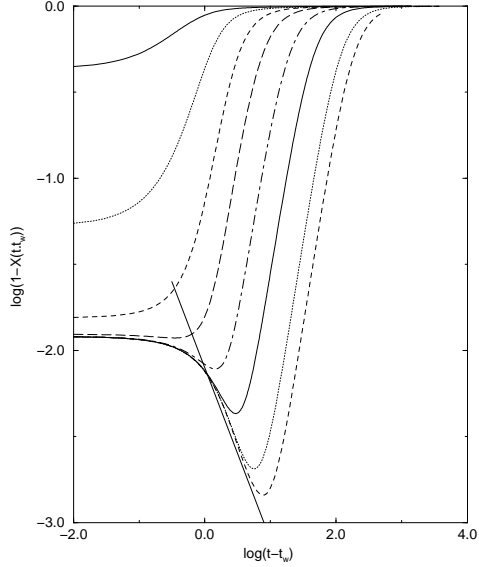}}
\end{center}
  \protect\caption[10]{1-$X(t,t_w)$ versus $t-t_w$. Different waiting
 times are shown.  From top to bottom
 $t_w$=1,3,10,30,100,300,1000,10000.  The straight continuous line shows
 $\frac{1}{t-t_w}$ behavior.  }
\end{figure}

We can also study the evolution of $X(t,t')$ as a function of
$C(t,t')$. Qualitatively we find very similar results 
in all dimensions:
in the $\beta$-regime $X$ is close to 1 while in the $\alpha$-regime it
decays very fast to zero. The 1-D and 3-D cases are depicted in figures
11 and 12 respectively.
Interestingly enough in the 3-D case we also find (figure 12) that all
the curves (except $t_w=1$ which is a too short time) cross at
$C^*=C_{norm}(t,t_w)\simeq 0.75$ (the value of $C^*$ increases for lower
temperatures) corresponding to a value of $X^*\simeq 0.25$ . The value
$C^*$ corresponds to the value of $C_{norm}(t_w,t)$ in the plateau in the
$\beta$-regime (see figure 7). It is possible to make all different
curves (corresponding to different values of $t_w$) collapse in a
universal one. We find that the scaling law,

\begin{equation}
X(t,t_w)=X((C^*-C_{norm}(t,t_w))\,t_w^{0.4})
\label{eq333}
\end{equation}

\noindent
for $C_{norm}(t_w,t)< 0.75$ nicely fits data for all different values of
$t_w$ (inset in figure 12). We can interpret naturally the value of $X$
in the two regimes as a ratio between the temperature of the system and
an effective temperature $T_f$. For $C_{norm}(t_w,t)>C^*$ we find
$X\simeq 1$ and the effective temperature of the system coincides with
the temperature of the thermal bath. In this $\beta$-regime the system
is in local equilibrium. For $C_{norm}(t_w,t)<C^*$ we find $X\simeq 0$
which means that the effective temperature is infinite. This is a
confirmation of the results obtained in section V.B interpreted within
the kinetic domain growth scenario. In that case it was found that the
effective temperature outside equilibrated domains was also
infinite. Note that the definition of an effective temperature is not
free of inconsistencies in the most general case and there is not
evidence {\em a priori} that both effective temperatures (in the
$\alpha$ regime) for the one time quantities and the effective
temperature obtained for the two time quantities coincide. A
result of this type was found in the Backgammon model for the
physical interpretation of the violation fluctuation dissipation ratio
\cite{BG2}.

\begin{figure}
\begin{center}
\leavevmode
\epsfysize=230pt{\epsffile{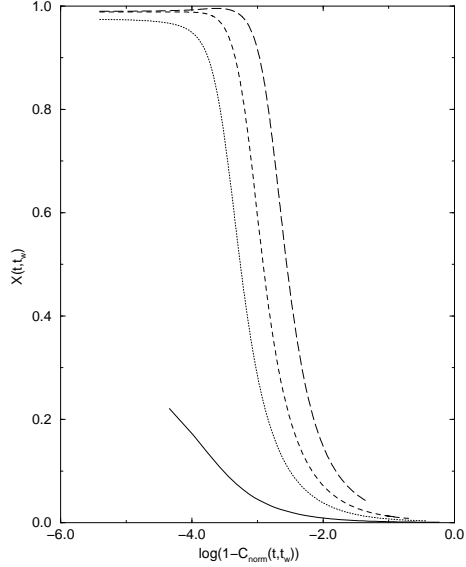}}
\end{center}
  \protect\caption[11]{
$X(t,t_w)$ versus $C_{norm}(t,t_w)$ in
1D at $T=0.0001$. Different waiting times are shown.
From top to bottom $t_w=1000,300,100,10$. 
}
\end{figure}

\begin{figure}
\begin{center}
\leavevmode
\epsfysize=230pt{\epsffile{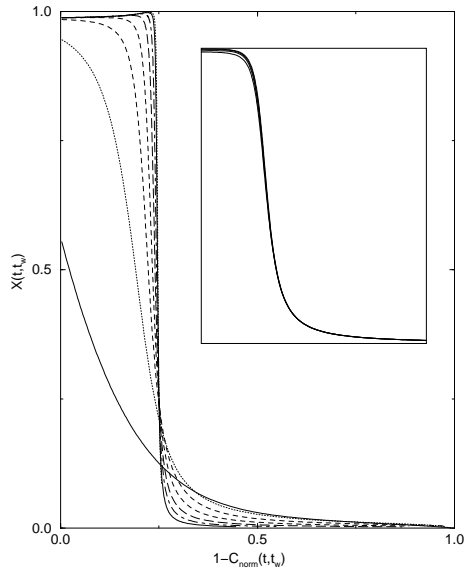}}
\end{center}
  \protect\caption[12]{
$X(t,t_w)$ versus $1-C_{norm}(t,t_w)$ in
3-D at $T=0.1$. Different waiting times are shown.
From bottom to top $t_w=1,3,10,30,100,300,1000,10000$.
In the inset we show the collapse of the data using the scaling relation
eq.(\ref{eq333})
for various values of $t_w=30,100,300,1000$. 
}
\end{figure}

\subsubsection{Correlation Between Replicas.}

Up to now we have found qualitatively similar results in the
off-equilibrium behavior in 1-D and 3-D. The question arises if it is
possible to infer from the dynamics the existence of a condensed growing
phase which could distinguish the 1-D from the 3-D behavior at finite
T. It has been recently suggested \cite{BBM} that it is possible to
characterize the dynamical behavior in terms of the overlap between two
replicas which start in the same initial configuration and follow
different noise realizations. We have analyzed the quantity,
$Q(t_w,t)-Q_{eq}$ where $Q_{eq}=\rho_{eq}^2$ for different values of
$t_w$.  In a disordered phase we expect that $Q(t_w,t)-Q_{eq}$
should decay to zero for long times because the two replicas depart one
from each other even if they are at the same initial condition at
$t_w$. This is the behavior we find in 1D (figure 13). In the 3-D case
in the condensed phase (see figure 14) the situation is different. Now
the two replicas remember they were in the same configuration at $t_w$
and do not depart indefinitely one from each other. The system is then
constrained to follow something similar to gutters or channels in phase
space \cite{BBM}. Intuitively it is not difficult to interpret this
result in terms of a domain growth process. We already now that the
dynamics of the LP model is essentially dominated by the growth of
correlated domains. In the 1-D case domains appear and disappear in time
because the system is in the disordered phase. In this case the typical
length of domains grow in time (we are at low temperatures) but domains
can always appear and disappear. The pattern structure of domains is in
some sense continuously changed in time. In the 3-D case, once domains
start to grow, they are not destroyed again. During the condensation
process the pattern structure of domains is essentially unchanged and it
is only rescaled in time.  Consequently the two-replicas overlap
$Q(t_w,t)-Q_{eq}$ does not decay to zero for long times.

\begin{figure}
\begin{center}
\leavevmode
\epsfysize=230pt{\epsffile{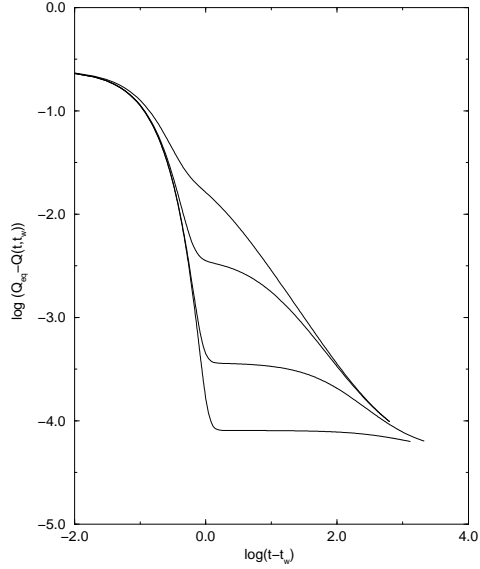}}
\end{center}
  \protect\caption[13]{
Correlation between replicas, $Q(t,t_w)-Q_{eq}$ against $t-t_w$
in 1D and $T=0.0001$. Different waiting times are shown.
From top to bottom $t_w=1$,10,100,1000.  
}
\end{figure}

\begin{figure}
\begin{center}
\leavevmode
\epsfysize=230pt{\epsffile{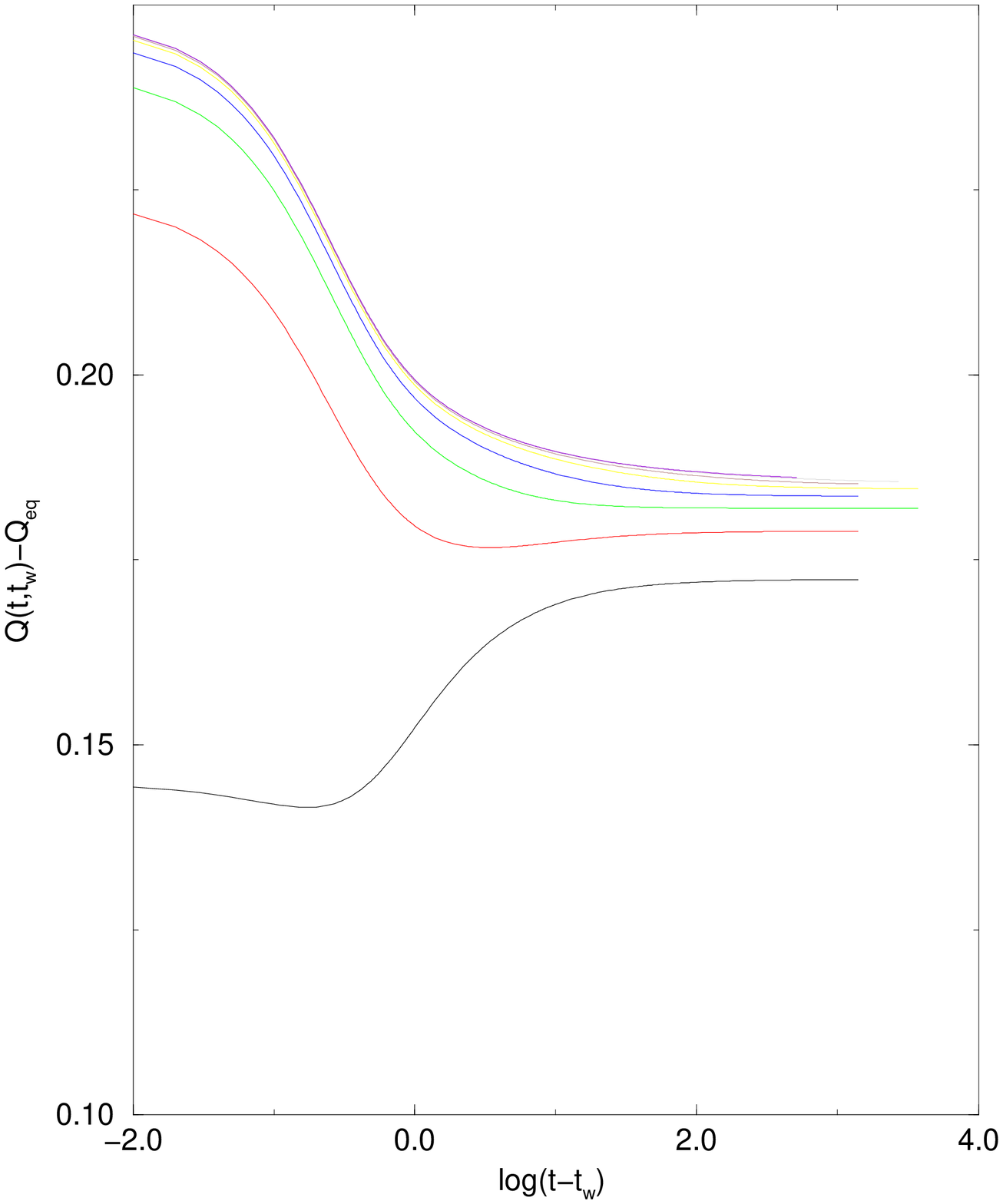}}
\end{center}
  \protect\caption[14]{ Correlation between replicas, $Q(t,t_w)-Q_{eq}$
against total menus waiting time 3-D and $T=0.1$. Different waiting
times are shown.  From bottom to top $t_w=1$,3,10,30,100,300,1000.  }
\end{figure}

\section{Conclusions}

We have analyzed in detail the dynamical properties of the LP model
originally introduced to understand the thermodynamic properties of hard
spheres lattice gases in the spherical approximation. From the viewpoint
of the dynamics this is an interesting model because it is exactly
solvable allowing a detailed investigation of the off-equilibrium
scenario. The model has no built-in disorder and slow dynamics appears
as a consequence of the short range dynamical constraints present in the
system. The dynamics of this model shares a large number of features
with the off-equilibrium dynamics of the spherical
Sherrington-Kirkpatrick model \cite{Kos} where dynamics is driven by the
macroscopic condensation of the system on the disordered ground
state. This gives support to the result that disorder is not an
essential ingredient for the dynamical glassy scenario \cite{BOME,MPR}
to be valid.

We have presented a detailed investigation of the dynamical equations of
the model by considering the one time and two time quantities. One
advantage of the model is that it includes short-range effects which are
not present in other type of mean-field models like the spherical SK
model \cite{Kos}.  If the initial configuration has density far from its
equilibrium value then there is a short-time regime where the system is
filled very fast in a spatially uncorrelated way. The typical time $t^*$
for this filling process is of order 1. It is only after that time that
slow relaxation starts when the system is spatially correlated and needs
to reorganize large regions in order to increase its density.

For values of time larger than $t^*$, we see that the dynamics of the
model shows striking similarities to the kinetics of domain growth of
ferromagnets with non conserved dynamics \cite{BRAY}. This is expected
since the dynamics proposed in eq.(\ref{eq5}) corresponds to the
coarsening dynamics in the so called model B in phase ordering
kinetics. In particular we have seen that for length scales smaller
than a typical correlation length $\xi$ (which grows in time) the
system is in local equilibrium and spatial fluctuations are determined
by the equilibrium temperature. Above the correlation length $\xi$ the
system is completely disordered, hence the typical temperature
associated to spatial fluctuations is infinite. The correlation length
$\xi$ is an accurate measure of the typical size of the growing
domains. It would be interesting to investigate the dynamics of the LP
model with the conserved dynamics of kinetic growth. In this case, we
expect that a similar scenario would apply but with different
exponents.

Further support to this domain growth scenario has been obtained from
the study of the two time quantities. In particular we find that
off-equilibrium relaxation proceeds in two well defined steps: a fast
$\beta$ relaxation process where the fluctuation dissipation relation is
obeyed, followed by a slow $\alpha$ process where time translation
invariance is lost and the fluctuation dissipation ratio is zero. The
physical meaning of the $\alpha$ and $\beta$ process is quite appealing
in terms of domain growth kinetics. The fast $\beta$-process is
associated to local rearrangements of densities inside domains whereas
the slow $\alpha$-process is associated to the growth of the typical
size of these domains. On the other hand, the response function shows no
aging and decays very fast to zero. This is the scenario of glassy
dynamics without anomaly in the response function typical of glassy
systems with short term memory. A study of the replica-replica overlap
has also revealed that the growth mechanism is different in the presence
or absence of condensation transition. In 1-D where there is no
condensation phase transition (the system is always in the disordered
phase) the domains appear and disappear in time even if their typical
size increases. In the 3-D case the domains always tend to grow and the
pattern structure of domains is always maintained. In this last case we
find that $\lim_{t_w\to\infty}\lim_{t\to\infty}Q(t_w,t)$ is finite
\cite{BBM}.

 We want to note that the relaxation of the density as well as
correlation functions do not display the phenomena of
stretching characteristic of glasses. This is due to the
oversimplification inherent to the model where only entropy barriers are
introduced through global constraints (while in the classical
hard-spheres model constraints are always local). It would be very
interesting to introduce some kind of local constraint which could
restore the main features observed in real glasses.  In this direction,
it would be quite interesting to extend this research by considering the
LP model with the dynamics recently proposed by Dean where local
densities can never become negative, a feature which is not considered
in the present dynamics \cite{DEAN}. This requires the introduction of
noise correlated with the local densities, a kind of local
constraint. Also it would be interesting if such a dynamics could be
exactly solved.

{\bf Acknowledgments}. We acknowledge Luis Bonilla, Enrique Diez and
Silvio Franz for stimulating discussions on this and related
subjects. F. G. P acknowledges to the Universiteit van Amsterdam for the
facilities during his stay there. The work by F.G.P. has been supported
by the DGCyT The work by F.R has been supported by
FOM under contract FOM-67596 (The Netherlands).

\vfill
\newpage


\end{document}